\documentclass[acp, manuscript]{copernicus}

\renewcommand{\thesubsection}{\thesection.\alph{subsection}}

\usepackage{latexsym,euscript,textcomp}
\usepackage{color,graphics,epsf,dcolumn}
\usepackage{epstopdf,ulem}
\usepackage{blindtext}
\usepackage{threeparttable}
\usepackage{xcolor}

\newcommand{\beq}{\begin{equation}}
\newcommand{\eeq}{\end{equation}}
\newcommand{\pt}{\partial}

\begin{document}
\nolinenumbers

\title{A critical analysis of the assumptions underlying the formulation of maximum potential intensity for tropical cyclones}


\Author[1,2]{Anastassia M.}{Makarieva}
\Author[1]{Andrei V.}{Nefiodov}

\affil[1]{Theoretical Physics Division, Petersburg Nuclear Physics Institute, Gatchina  188300, St.~Petersburg, Russia}
\affil[2]{Institute for Advanced Study, Technical University of Munich,  Garching 85748, Germany}


\runningtitle{A critical analysis of the assumptions}

\runningauthor{Makarieva et al.}

\correspondence{A. M. Makarieva (ammakarieva@gmail.com)}

\received{}
\pubdiscuss{} 
\revised{}
\accepted{}
\published{}


\firstpage{1}

\maketitle

\begin{abstract}
Emanuel{\textquoteright}s concept of maximum potential intensity (E-PI) estimates the maximum velocity of tropical cyclones from environmental parameters
assuming thermal wind (gradient-wind and hydrostatic balances) and slantwise neutrality in the free troposphere. E-PI{\textquoteright}s key equation relates proportionally the  radial gradients of saturated moist entropy and angular momentum. Here the E-PI derivation is reconsidered to show that the thermal wind and slantwise neutrality imply zero radial gradients of saturation entropy and angular momentum at an altitude where, for a given radius, the tangential wind has a maximum.  It is further shown that, while E-PI{\textquoteright}s key equation requires that, at the point of maximum tangential wind, the air temperature must increase towards the storm center, the thermal wind equation dictates the opposite. From the analysis of the equations of motion at the altitude of maximum tangential wind in the free troposphere, it is concluded that here the air flow must be supergradient. This implies that the supergradiency factor (a measure of the gradient-wind imbalance) must change in the free troposphere as the air flow tends to restore the balance. It is shown that such a change modifies  the derivative of saturation entropy over angular momentum, which cannot therefore remain constant in the free troposphere as E-PI requires.  The implications of these findings for the internal coherence of E-PI, including its boundary layer closure, are discussed.
\end{abstract}

\introduction  
\label{intr}

Tropical storms threaten human lives and livelihoods. Numerical models can simulate a wide range of storm intensities under the same environmental conditions \citep[e.g.,][]{tao20}. Thus it is desirable to  have a reliable theoretical framework that would, from the first principles, confine model outputs to the domain of reality \citep{emanuel20}.  The theoretical formulation for maximum potential intensity (E-PI) of tropical cyclones by \citet{em86} has been long considered as an approximate upper limit on storm intensity (see discussions by \citet{garner15}, \citet{kieu2016} and \citet[][]{kowaleski16}). Studies have shown that the maximum wind (observed or modelled) can be larger than E-PI due to  supergradient wind ({\textquotedblleft}superintensity{\textquotedblright}) \citep[e.g.,][]{persing2003,montgomery06,bryan09b,em19,Li20c}. Here we re-consider the assumptions behind E-PI to show that they are mutually incompatible at the point of maximum wind.

The E-PI formulation is based on the thermal wind equation and the assumption of slantwise neutrality in the free troposphere.
In Section~\ref{der} we repeat the E-PI derivation following \citet{em86} but focusing on the altitude where tangential velocity has a local maximum $\pt v/\pt z = 0$. We show that here E-PI predicts zero radial gradients of saturation entropy $s^*$ and angular momentum thus not permitting non-trivial solutions and being inapplicable for the assessment of maximum winds. 

For readers immediately interested in the underlying physics, here is a brief explanation. The gradient-wind balance consists
in the equality of the centripetal and centrifugal forces: the radial pressure gradient per unit density and the squared tangential velocity divided by radius.
Where $\pt v/\pt z = 0$, the latter force is invariant over $z$. For the thermal wind equation to apply, the gradient wind (determined by the radial pressure gradient per unit air density) must also be constant over $z$. In hydrostatic equilibrium, this is the case when the radial and vertical gradients of temperature $T$ over pressure $p$ are equal (see appendix~\ref{prgr}).  When $\pt s^*/\pt z = \pt s^*/\pt r = 0$, this condition is fulfilled: the temperature gradients in both directions are moist adiabatic. In real cyclones, the radial pressure gradient  diminishes with height changing its sign in the mid troposphere. The thermal wind equation at the point $\pt v/\pt z = 0$ cannot hold. Another perspective on the same problem is that E-PI constrains the slope of angular momentum surfaces, and this predicted slope is never zero -- although it must be so where $\pt v/\pt z = 0$ (see Section~\ref{bound}).

In Section~\ref{key} we discuss why the incompatibility between E-PI{\textquoteright}s assumptions is not explicit in the resulting E-PI formula.  In Section~\ref{tgr} we show how this incompatibility can be explicated by combining the E-PI formula with the definition of saturated moist entropy. This reveals that the E-PI formula and the thermal wind equation from which it derives, predict the opposite signs of the radial temperature gradient at the point of maximum tangential wind. In Section~\ref{mot} we discuss additional dynamic constraints on E-PI from the equations of motion. In Section~\ref{bound}  we discuss the implications of our findings for the boundary layer closure in E-PI. In the view of the obtained results, the concluding Section~\ref{dis} discusses the general coherence of E-PI and some issues with its verification by numerical modelling. 

\section{E-PI derivation for the point where $\pt v/\pt z = 0$}
\label{der}

We begin with the consideration of \citeauthor{em86}{\textquoteright}s~\citeyearpar{em86} Eqs.~(1)--(8), which describe a steady axisymmetric hurricane vortex. Absolute angular momentum per unit mass is
\begin{equation}\label{M}
M \equiv  v r + \frac{1}{2}fr^2,
\end{equation}
where $v$ is the tangential velocity and $r$ is the distance from the cyclone center.
The Coriolis parameter $f \equiv 2 \Omega \sin \varphi$ is assumed constant ($\varphi$ is latitude, $\Omega$ is the angular velocity of Earth{\textquoteright}s rotation).

With $g$ as the acceleration of gravity and $\alpha \equiv 1/\rho$ as the specific volume, the hydrostatic balance is
\beq\label{he1}
\alpha \frac{\pt p}{\pt z} = - g.
\eeq
The radial balance of forces can be written as
\beq\label{gwb1}
b \alpha \frac{\pt p}{\pt r} = \frac{v^2}{r} + fv = \frac{M^2}{r^3} - \frac{1}{4} f^2 r.
\eeq
Here we have introduced the \textit{supergradiency} factor $b$. It defines the degree to which the flow is radially unbalanced: $b > 1$ for the supergradient flow when 
the outward-pushing centrifugal force is larger than the inward-pulling pressure gradient.
\citet[][his Eq.~(3)]{em86} assumed gradient-wind balance, i.e., $b = 1$. 

Using Eq.~\eqref{he1}, Eq.~\eqref{gwb1} can be written as follows
\beq\label{gwb2}
g \left(\frac{\pt z}{\pt r}\right)_p = - g \left(\frac{\pt p}{\pt r}\right)_z  \left(\frac{\pt p}{\pt z}\right)_r^{-1}  = 
\frac{1}{b} \left(\frac{M^2}{r^3} - \frac{1}{4} f^2 r\right).
\eeq

Hydrostatic equilibrium \eqref{he1} is rewritten as
\beq\label{he2}
g \left(\frac{\pt z}{\pt p}\right)_r = g \left(\frac{\pt p}{\pt z}\right)_r^{-1}= -\alpha.
\eeq

Taking the derivative of Eq.~\eqref{gwb2} with respect to $p$ at constant $r$, and of Eq.~\eqref{he2} with respect to $r$ at constant $p$,
we obtain
\beq\label{em6}
 \frac{1}{r^3}\left(\frac{\pt M^2}{\pt p} \right)_r - \left(\frac{M^2}{r^3} - \frac{1}{4} f^2 r\right)\frac{1}{b} \left(\frac{\pt b}{\pt p}\right)_r  = -b \left(\frac{\pt \alpha}{\pt r}\right)_p .
\eeq
In gradient-wind balance, i.e., with $b = 1$ and $(\pt b /\pt p)_r = 0$, the second term on the left-hand side of Eq.~\eqref{em6} is zero, and it becomes \citeauthor{em86}{\textquoteright}s~\citeyearpar{em86}  thermal wind equation.

\citet{em86} assumed reversible thermodynamics neglecting the condensate loading.\footnote{\citet{mpi4-jas} estimated that the effect of condensate loading on E-PI{\textquoteright}s formulation is minor.} In this case $\alpha$ is a function of $p$ and saturation entropy $s^*$
alone and
\beq\label{al}
\left(\frac{\pt \alpha}{\pt r}\right)_p = \left(\frac{\pt \alpha}{\pt s^*}\right)_p \left(\frac{\pt s^*}{\pt r}\right)_p.
\eeq
For the definition of $s^*$ see Eq.~(\ref{s}) in appendix~\ref{altf}.

Using the first law of thermodynamics, it can be shown that $(\pt \alpha/\pt s^*)_p = (\pt T/\pt p)_{s^*}>0$, where
the last expression is the moist adiabatic temperature gradient. Then Eq.~\eqref{em6} becomes
\beq\label{em8}
\frac{1}{r^3}\left(\frac{\pt M^2}{\pt p} \right)_r - \left(\frac{M^2}{r^3} - \frac{1}{4} f^2 r\right)\frac{1}{b} \left(\frac{\pt b}{\pt p}\right)_r  = 
- b\left(\frac{\pt T}{\pt p}\right)_{s^*} \left(\frac{\pt s^*}{\pt r}\right)_p.
\eeq

At this point \citet{em86} involved the assumption of slantwise neutrality in the form $s^* = s^*(M)$, i.e., assuming that 
saturation entropy is a function of angular momentum alone. We will now consider the implications of
this assumption applied at the point where $\pt v/\pt z = 0$ together with Eq.~\eqref{em8} where $(\pt b/\pt p)_r = 0$.

Whenever $\pt v/\pt z = 0$, we have $\pt M/\pt z = 0$ and $(\pt M/\pt p)_r = 0$ and, by consequence from Eq.~\eqref{em8}, $(\pt s^*/\pt r)_p = 0$.
On the other hand, since $s^* = s^*(M)$ and $(\pt s^*/\pt p)_r = (ds^*/dM) (\pt M/\pt p)_r$, we also have $(\pt s^*/\pt p)_r = 0$
(excluding the unrealistic case $ds^*/dM = 0$). But since $(\pt s^*/\pt p)_r = 0$ and 
$(\pt s^*/\pt r)_p = 0$, this means that whenever $\pt v/\pt z = 0$, the radial gradient of saturation entropy is zero: $\pt s^*/\pt r = 0$.
Furthermore, since $\pt s^*/\pt r = (ds^*/dM) \pt M/\pt r$, the
radial gradient of angular momentum is also zero: $\pt M/\pt r = 0$. These conclusions 
do not depend on the value of $b$.

In real cyclones, saturation entropy increases towards the storm center in the lower atmosphere: $\pt s^*/\pt r < 0$ \citep[e.g.,][their Fig.~4f]{montgomery06}. Equation~\eqref{em8} applied for $\pt v/\pt z = 0$ makes it clear that for $\pt s^*/\pt r < 0$ the supergradiency factor $b$
must grow with altitude (i.e., $(\pt b/\pt p)_r < 0$):
\beq\label{dbz}
\left(\frac{\pt T}{\pt p}\right)_{s^*} \left(\frac{\pt s^*}{\pt r}\right)_p = \left(\frac{M^2}{r^3} - \frac{1}{4} f^2 r\right)\frac{1}{b^2} \left(\frac{\pt b}{\pt p}\right)_r .
\eeq   
As a side note, the supergradiency factor $b$ can change due to thermodynamics (which controls the pressure gradient) and dynamics (which controls the centrifugal force), see Eq.~\eqref{gwb1}. We show in appendix~\ref{prgr} that the thermodynamic change of $b$ over $z$ in the boundary layer is small, not exceeding a few per cent over $1$~km. Any significant deviation from the gradient-wind balance should be due to turbulent friction.

Our conclusion so far is that the thermal wind equation and the assumption of slantwise neutrality are incompatible with $\pt v/\pt z = 0$.

\section{E-PI{\textquoteright}s key relationship}
\label{key}

We will now see why this incompatibility is not explicit in the resulting E-PI formula. We put  $(\pt b/\pt p)_r = 0$ in Eq.~\eqref{em8}. With $b=1$, the four equations below correspond to \citeauthor{em86}{\textquoteright}s~\citeyearpar{em86}  Eqs.~(10)--(13).

Using $(\pt s^*/\pt r)_p = (ds^*/dM) (\pt M/\pt r)_p$, \citet{em86}
divided Eq.~\eqref{em8} by $(\pt M/\pt r)_p$ to obtain
\beq\label{em10}
\left(\frac{\pt r}{\pt p}\right)_M =  \frac{br^3}{2M} \frac{ds^*}{dM} \left(\frac{\pt T}{\pt p}\right)_{s^*}.
\eeq
This equation was integrated along $M$ and $s^*=s^*(M)$ surfaces. The result is 
\beq\label{em11}
\frac{1}{r^2} - \frac{1}{r_o^2} = -\frac{b(T-T_o)}{M}\frac{ds^*}{dM},
\eeq
where $T_o$ is the temperature at $r=r_o$. The transition from Eq.~\eqref{em8} to Eq.~\eqref{em10} requires $s^*=s^*(M)$ locally. The transition from Eq.~\eqref{em10} to Eq.~\eqref{em11} requires additionally that $ds^*/dM$ is constant over $r$ following $M$ or $s^{*}$ surface.

For $r_o \gg r$, Eq.~\eqref{em11} takes the form 
\beq\label{em12}
-r^2 b (T-T_o)\frac{ds^*}{dM}  = M.
\eeq

Finally, Eq.~\eqref{em12} is multiplied by $\pt M/\pt r$ to yield  \citeauthor{em86}{\textquoteright}s~\citeyearpar{em86}  Eq.~(13),
which, with $b = 1$, is E-PI{\textquoteright}s key equation:
\beq\label{em13}
-bT \varepsilon \frac{\pt s^*}{\pt r}  = \frac{M}{r^2} \frac{\pt M}{\pt r}.
\eeq
Here $\varepsilon \equiv (T - T_o)/T$ can be interpreted as Carnot efficiency; $T_o$ can be a function of $b$. 
\citet[][their Eqs.~(35) and (36)]{mpi4-jas} obtained this result, i.e.,  Eq.~\eqref{em13} with $b \ne 1$, using a different approach. This result can also be obtained from  \citeauthor{bryan09b}{\textquoteright}s~\citeyearpar{bryan09b}  Eq.~(20) using $u=0$ \citep[see Section~\ref{mot} below and appendix C of][]{mn21}. 
With the gradient-wind assumption relaxed, 
$\varepsilon$ is replaced with $b \varepsilon$, which is greater than Carnot efficiency for supergradient storms.

These derivations, of \citet{bryan09b}, \citet{mpi4-jas}, and Eqs.~\eqref{em10}--\eqref{em13} with a constant $b\ne 1$, assume that in the free troposphere the air motion conserves not only the angular momentum, but also the supergradiency factor $b \ne 1$. Such motion, while mathematically possible, is not physically plausible: in the real free troposphere the flow will tend  to restore the gradient-wind balance, i.e., $b \ne 1$ will change to $b \simeq 1$ (with a minor deviation from unity determined by how small the turbulent friction is). If, as it enters the free troposphere, the air flow is supergradient with $b > 1$, then, as it begins to relax to gradient balance, $(\pt b/\pt p)_r > 0$ in Eq.~\eqref{em8}. The absolute magnitude of $ds^*/dM$ retrieved from Eq.~\eqref{em10} is then smaller than it is when $(\pt b/\pt p)_r = 0$. This indicates that $|ds^*/dM|$ should increase in the upper troposhere where the air reaches gradient-wind balance ($b \simeq 1$ and $(\pt b/\pt p)_r \simeq 0$).  

Focusing on the point where $\pt v/\pt z = 0$, we notice that E-PI{\textquoteright}s key equation was obtained by dividing Eq.~\eqref{em8} by $(\pt M/\pt r)_p = 0$, integrating the resulting equation along $M$ surface, and multiplying it again by $\pt M/\pt r = 0$. In the resulting formula, after this dividing and multiplying by zero, the inapplicability of E-PI to the point where $\pt v/\pt z = 0$ became implicit. However, as we show in the next section, it can be explicated at the point of maximum tangential wind.

\section{Radial temperature gradient at the point of maximum tangential wind}
\label{tgr}
We will now show that, at the  point of maximum tangential wind, where $\pt v/\pt r = \pt v/\pt z = 0$,
E-PI{\textquoteright}s key equation and the thermal wind equation, from which the former is derived,
predict the opposite signs for the radial temperature gradient. 

\subsection{Constraints on $\pt T/\pt r$ from the definition of saturation entropy}
\label{constr0}

Since saturation entropy $s^*$ is a state variable, its radial gradient can be expressed in terms of the radial gradients of air pressure and temperature (see Eq.~\eqref{Tds3}):
\begin{gather} 
\frac{T}{1+\zeta} \frac{\partial s^*}{\partial r} = -\alpha_d  \mathcal{C} \frac{\partial p}{\partial r}, \label{sr}\\
\mathcal{C} \equiv 1-  \frac{1}{\Gamma}\left( \frac{\partial T}{\partial r}\right)  \left( \frac{\partial p}{\partial r}\right)^{-1} =
1 -\frac{1}{\Gamma}\left( \frac{\partial T}{\partial p}\right)_z     ,  \label{C}
\end{gather}
where $\zeta \equiv L \gamma_d^*/(R T)$, $R = 8.3$~J~mol$^{-1}$~K$^{-1}$ is the universal gas constant, $\gamma_d^* \equiv p_v^*/p_d$, $p_v^*$ is the  partial pressure of saturated water vapor, $p_d$ is the partial pressure of dry air, $L \simeq 45$~kJ~mol$^{-1}$ is the latent heat of vaporization, $\Gamma$ (K Pa$^{-1}$) is the moist adiabatic lapse rate of air temperature (see its definition \eqref{Gm}), and $\alpha_d \equiv 1/\rho_d$ is the inverse dry air density. Below we assume $\alpha_d \simeq \alpha$, where $\alpha \equiv 1/\rho$ is the inverse air density. Equation (\ref{sr}) does not contain any assumptions but follows directly from the definition of saturation entropy. 

Where $\partial v/\partial r = 0$, we have $\partial M/\partial r = v + fr$ and
\noindent
\begin{equation}\label{dpr}
b\alpha \frac{\partial p}{\partial r} = \frac{v}{r} \frac{\partial M}{\partial r} = \frac{M}{r^2} \frac{\partial M}{\partial r}.
\end{equation}
The last equality assumes $v \gg fr/2$. Combining Eqs.~(\ref{sr})--(\ref{dpr})  we obtain an alternative version of Eq.~\eqref{em13},
\begin{equation} 
\label{alt1}
-\frac{bT}{\mathcal{C}(1+\zeta)} \frac{\partial s^*}{\partial r} =   \frac{M}{r^2}\frac{\partial M}{\partial r}, \quad  \frac{\pt v}{\pt r} = 0. 
\end{equation}
Equation~\eqref{em13}  assumes thermal wind and slantwise neutrality, Eq.~(\ref{alt1}) assumes $\pt v/\pt r = 0$.
Therefore, where $\pt v/\pt r = 0$, Eq.~(\ref{alt1}) is more general than Eq.~\eqref{em13} and must hold simultaneously with the latter.

Combining Eqs.~\eqref{em13} and \eqref{alt1} and using the definition of $\mathcal{C}$ we obtain
\begin{gather} 
\mathcal{C} = \frac{1}{\varepsilon (1+ \zeta)}, \label{comp1}\\ 
\frac{\pt p}{\pt r} =  \frac{1}{1 - \mathcal{C}}  \frac{1}{\Gamma}\frac{\pt T}{\pt r}, \quad \frac{\pt v}{\pt r} = 0.  \label{comp2}
\end{gather}
 
This shows that a E-PI cyclone would only be possible in the presence of a non-zero radial gradient of air temperature at the point of maximum tangential wind\footnote{Equations~\eqref{comp1} and \eqref{comp2} clarify why hypercanes cannot exist. With $\varepsilon(1+\zeta)\to 1$, $\mathcal{C} \to 1$ and $\pt T/\pt r \to 0$.
The radial pressure gradient in Eq.~\eqref{comp2} is then undefined rather than infinite \citep[cf.][]{em88}, see also \citet[][appendix~B]{mn21}.}.
\citet{mpi4-jas} obtained this result for dry E-PI cyclones. 

The maximum Carnot efficiency estimated from temperatures $T_o$ and $T=T_b$ observed, respectively, in the outflow and at the top of the boundary layer,  is $\varepsilon= 0.35$ \citep{demaria94}. Assuming that $T_b$ does not usually exceed $303$~K ($30${\textcelsius}), the minimum value of $(1+\zeta)^{-1}\simeq  0.5$ is larger than $\varepsilon$. It corresponds to the largest $\gamma_d^* \simeq 0.05$ for $T_b = 303$~K  and $p_d \simeq p = 800$~hPa. The  partial pressure $p_v^*$ of saturated water vapor  and, hence, $\gamma_d^*$ depend exponentially on air temperature. The realistic temperatures at the top of the boundary layer are commonly significantly lower than $303$~K. 

\begin{figure*}[tbp]
\begin{minipage}[p]{0.50\textwidth}
\centering\includegraphics[width=1\textwidth,angle=0,clip]{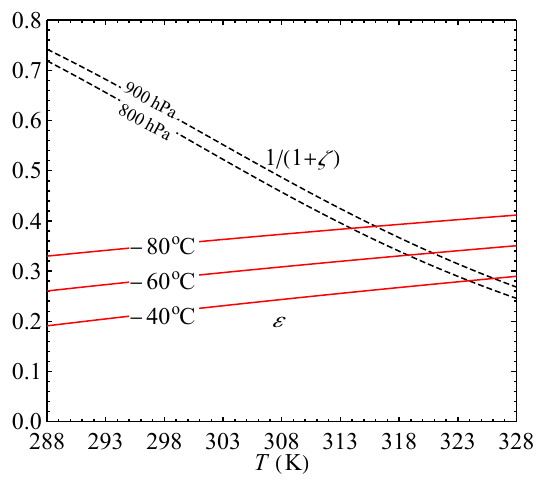}
\end{minipage}
\caption{Parameters $\varepsilon \equiv (T - T_o)/T$ versus $1/(1+\zeta)$ as dependent on temperature $T$; $\varepsilon$ curves correspond to different outflow temperatures $T_o$;  $1/(1+\zeta)$ curves correspond to $p_d$ values of  $800$ and $900$~hPa, see Eq.~(\ref{Gm}). }
\label{fig1}
\end{figure*}

Thus, under observed atmospheric conditions $[\varepsilon(1+\zeta)]^{-1} > 1$ (Fig.~\ref{fig1}). This means that for the E-PI cyclone to exist, i.e.,  for $\pt p/\pt r > 0$, the air temperature must grow in the direction of the cyclone center, i.e.,  $\pt T/\pt r < 0$ where $\pt v/\pt r = 0$. We emphasize that, to be valid, E-PI requires  a specific value of $\pt T/\pt r <0$ as determined by Eq.~\eqref{comp1}. At observed temperatures, E-PI requires $\mathcal{C} \simeq 2$ (i.e., $(\pt T/\pt p)_{z} \simeq - \Gamma$). Notably, due to the change of the sign of the last term in Eq.~\eqref{comp2} at high temperatures, see Fig.~\ref{fig1}, E-PI requires that at high temperatures the air temperature at the point of maximum tangential wind should decline in the direction of the storm center, i.e.,  $\mathcal{C} < 1$.

\subsection{Thermal wind constraints on $\pt T/\pt r$}
\label{constr}

We will now show that the thermal wind equation does not permit $\pt T/\pt r < 0$ where $\pt v/\pt z = 0$. 
Hydrostatic balance~\eqref{he1} can be cast into the following form:
\beq\label{he}
\frac{\pt \ln p}{\pt z} = -\frac{1}{h}, \quad h \equiv \frac{RT}{g M_a},
\eeq
where $M_a$ is air molar mass (here assumed to be constant as in E-PI) and $R$ is the universal gas constant.

Where $\pt v/\pt z = 0$, the derivative with respect to $z$ of the right-hand side of Eq.~\eqref{gwb1} is zero, so
 taking into account that  $\alpha = gh/p$ according to Eq.~\eqref{he}, we have
\beq\label{dz}
\frac{\pt b}{\pt z} h \frac{\partial \ln p}{\partial r} + b \frac{\pt h}{\pt z} \frac{\pt \ln p}{\pt r}
+ b h \frac{\pt}{\pt z} \left(\frac{\pt \ln p}{\pt r}\right) = 0,\quad \frac{\pt v}{\pt z} = 0.
\eeq
Interchanging the order of differentiation in the last term in Eq.~\eqref{dz}, using Eq.~\eqref{he}  and noting that $d \ln T = d \ln h$, we obtain 
\beq\label{dz1}
\frac{\partial \ln p}{\partial r} \left( \frac{\pt \ln b}{\pt z} + \frac{\pt \ln T}{\pt z} \right) =  -\frac{1}{h} \frac{\pt \ln T}{\pt r},\quad \frac{\pt v}{\pt z} = 0. 
\eeq

If the gradient-wind balance holds in the free troposphere, we have $\pt b/\pt z = 0$. 
Using the definition of $h$ \eqref{he}, Eq.~\eqref{dz1} then becomes, cf. Eq.~\eqref{comp2}:
\beq\label{comp3}
\frac{\pt p}{\pt r} = - \rho g \frac{\pt T}{\pt r}  \left(\frac{\pt T}{\pt z} \right)^{-1} , \quad \frac{\pt v}{\pt z} = 0.
\eeq
Since $\pt T/\pt z < 0$, for there to be a cyclone, i.e.,  for $\pt p/\pt r > 0$, the air temperature must locally decline
in the direction of the cyclone center, i.e.,  $\pt T/\pt r > 0$ where $\pt v/\pt z = 0$. 

This result can be directly derived from the thermal wind equation, Eq.~(6) of \citet{em86}
and Eq.~(5) of \citet{emanuel11}, see also our Eq.~\eqref{em6}. It says that where the balanced wind is maximum over $z$, $(\pt M/\pt p)_r = 0$, we have 
$(\pt \alpha/\pt r)_p = 0$ and, hence, $(\pt T/\pt r)_p = 0$.  In the boundary layer of tropical cyclones, the isobars rise outward from the center, $(\pt z/\pt r)_p > 0$. 
With $\pt T/\pt z < 0$, the coincidence of isobars and isotherms means that $\pt T/\pt r > 0$.

\section{Constraints from the equations of motion}
\label{mot}

For the condition $\pt T/\pt r < 0$ to hold, it follows from Eq.~\eqref{dz1} that the supergradiency  factor $b$ must increase with altitude  at the point of maximum tangential wind, $\pt \ln b/\pt z > -\pt \ln T/\pt z > 0$. We will now show that $b= 1$ is incompatible with $\pt b/\pt z \ne 0$ where $\pt v/\pt z = 0$. It is not possible to retain the gradient-wind balance assumption locally but to relax it in the vicinity of this point.

The steady-state equations of axisymmetric motion for a frictionless atmosphere can be written as
\begin{subequations}
\begin{align}\label{m1}
\alpha \frac{\partial p}{\partial r} &= \frac{v^2}{r} + fv - u\frac{\pt u}{\pt r}-w\frac{\pt u}{\pt z}, \\ \label{m2}
0 &=  \frac{u}{r} \frac{\partial (r v)}{\partial r} +f u + w \frac{\pt v}{\pt z}, \\ \label{m3}
\alpha \frac{\partial p}{\partial z} &= - g - u\frac{\pt w}{\pt r}-w\frac{\pt w}{\pt z}, 
\end{align}
\end{subequations}
where $u$ is the radial velocity and $w$ is the vertical velocity.

Where $\pt v/\pt z = 0$, the derivative of the right-hand part of Eq.~\eqref{gwb1} with respect to $z$ is zero, so we have from Eq.~\eqref{m1}
\beq\label{bz}
\alpha \frac{\partial p}{\partial r} \frac{\pt \ln b}{\pt z} = -\frac{\pt}{\pt z} \left(\alpha \frac{\partial p}{\partial r}\right) =
\frac{\pt u}{\pt z} \frac{\pt u}{\pt r} + u \frac{\pt^2 u}{\pt z\pt r} + \frac{\pt w}{\pt z} \frac{\pt u}{\pt z} + w \frac{\pt^2 u}{\pt z^2}.
\eeq

For $\pt v/\pt z = 0$ we have from Eq.~\eqref{m2} that $u = 0$.
If $b = 1$ in Eq.~\eqref{gwb1}, the sum of the last two terms in Eq.~\eqref{m1} is zero. For $u=0$ and $w \ne 0$ (the eyewall)\footnote{\citet[][p.~3054]{bryan09b} noted that in numerical simulations the point of maximum tangential wind often coincides with the point of maximum vertical wind and that 
{\textquotedblleft}numerical simulations and observations often show that $u \approx 0$ at the location of maximum tangential velocity{\textquotedblright}.}, this means that $\pt u/\pt z = 0$. With $u = 0$ and $\pt u/\pt z = 0$, the first three terms in the right-hand part of Eq.~\eqref{bz} are zero.

The radial velocity $u$ changes its sign at the point of maximum tangential wind, where $u = 0$.
Below this point, there is convergence and $u< 0$, while above this point there is divergence and $u > 0$.
Usually the horizontal level that separates $u < 0$ and $u > 0$ is close to the top of the boundary layer, see, e.g., 
\citeauthor{bryan09b}{\textquoteright}s~\citeyearpar{bryan09b}
Fig. 11 for modelling and \citeauthor{montgomery06}{\textquoteright}s~\citeyearpar{montgomery06}  Fig.~4b for real cyclones.
From the conditions that $\pt u/\pt z = 0$ at the point where $\pt v/\pt z = 0$ and $\pt u/\pt z \ge 0$ in the vicinity of this point,\footnote{In mathematical analysis, this point is called \textit{a stationary point of inflection}, or \textit{saddle point}. It 
is the point on a curve at which the curvature changes sign.}
it follows that the second derivative of $u$ with respect to $z$ is zero at the point where $\pt v/\pt z = 0$. Then the fourth term in the right-hand part of Eq.~\eqref{bz} is zero as well. This means that $\pt b/\pt z = 0$, if $b = 1$ where $\pt v/\pt z = 0$. This shows that it is generally not possible to specify $b$ and $\pt b/\pt z$ independently.\footnote{\citet{smith08} brought up a related argument. They stated that, with $b=1$ at the top of boundary layer, E-PI {\it implicitly} assumed gradient-wind balance within the boundary layer.  \citet[][p.~2239]{emanuel11} replied that the boundary layer closure in E-PI did not require such an {\it explicit} assumption. \citet[][p.~553]{smith08} were correct for their particular model of the boundary layer, which assumed  $\pt u/\pt z =\pt v/\pt z =0$. In this case Eq.~\eqref{bz} yields $\pt b/\pt z = 0$ from $b=1$.}

In the eyewall with $\pt u/\pt z >0$ and $w > 0$, it follows from Eqs.~\eqref{m1} and \eqref{m2} that $b > 1$ for $\pt v/\pt z = 0$. In other words, those tropical cyclones that  have their maximum wind in the free troposphere, must be supergradient (cf. Eq.~\eqref{dbz}).
The conventional \textit{balanced} E-PI, which assumes $b=1$ in the free troposphere, has no solutions under observed atmospheric conditions.
With $b \ne 1$ unknown, E-PI is not a closed theory.

\section{Implications for the boundary layer closure in E-PI}
\label{bound}

We are now in a position to discuss where $\pt v/\pt z = 0$ is realized in real cyclones and in models. Surprisingly, despite all the research emphasis on {\it maximum potential intensity}, the question of where this maximum is located along the vertical axis does not appear to have received consistent attention: observational studies of vertical $v$ profiles exclude the boundary layer (see below).
For case studies, \citet[][their Fig.~4a]{montgomery06} reported that, for Hurricane Isabel (2003), the mean tangential wind in the eyewall (40~km~$\le r \le 50$~km)  has a maximum at a height of about $1$~km, where it is approximately $50\%$ greater than its surface value of approximately $50$~m~s$^{-1}$. 
Hurricanes Ivan (2004), Wilma (2005), Frances (2004), Helene (2006), and Dennis (2005), as shown, respectively, in Figs.~1c, 7b, 7c and 7d of \citet{stern2014} and Fig.~5a of \citet{stern2009}, display the same feature: for $r \gtrsim 40$~km, the slopes of $v$ contours change sign, thus indicating a maximum of $v$, at or below $1$~km. \citeauthor{peng2018}{\textquoteright}s~\citeyearpar{peng2018} Figs.~4 and 5 likewise show a maximum of tangential wind at $\sim 0.5$~km altitude for their simulated cyclones. Thus, the increase of tangential wind with altitude with a maximum near the top of the boundary layer in the eyewall appears to be a common feature in real storms as well as in models. 

In E-PI, \citeauthor{em86}{\textquoteright}s~\citeyearpar{em86} Fig.~1 presents a scheme of the boundary layer with constant $s^*$ surfaces that are vertical and constant $M$ surfaces that are not vertical,  but become approximately so at the boundary layer top. This scheme is consistent with the assumption \citet[][p.~3045]{bryan09b} assigned to E-PI,  that the maximum gradient wind {\textquotedblleft}is located at the top of the boundary layer{\textquotedblright} where {\textquotedblleft}viscous terms become negligible{\textquotedblright}. \citet[][p.~2239]{emanuel11} explained that for the boundary layer closure in E-PI it is sufficient that {\textquotedblleft}entropy is well mixed along angular momentum surfaces, which are approximately vertical in the boundary layer{\textquotedblright}, and vertical is how \citet{peng2018} show these surfaces in their Fig.~10b for E-PI. \citet[][p.~553]{smith08} also interpreted E-PI{\textquoteright}s boundary layer closure as presuming $\pt v/\pt z = 0$ within the boundary layer. Approximately vertical $v$ contours near the radius of maximum wind can be observed in real storms as well \citep[see, e.g.,][their~Fig.~1a,b for Hurricane Ivan (2004)]{stern2014}.

In  \citeauthor{bryan09b}{\textquoteright}s~\citeyearpar{bryan09b} control simulation, tangential
velocity in the eyewall increases with altitude within the lower $1$~km (see their Fig. 4b).
According to \citet[][p.~3050]{bryan09b}, the assumption that the maximum tangential velocity is achieved
 at the top of the boundary layer, {\textquotedblleft}is needed to match the free-atmosphere component to
the boundary layer closure in E-PI{\textquotedblright}.  In some discrepancy with this interpretation,
\citet[][their Fig.~5a,c]{stern2011} indicated  that E-PI, rather, presumes that the maximum tangential wind is
located at the surface $z = 0$, where $\pt v/\pt z < 0$, and monotonously declines with height.\footnote{\citet{stern2011} did not verify this pattern from observations, as they confined their consideration to above $2$~km. Subsequent studies retained this limitation 
\citep{hazelton2013,stern2014}; the recent study of \citet{fischer2022} does not show the lower $2$~km
in their Fig.~8 for the vertical profiles of tangential wind -- despite the data are available down to the lowest $500$~m.}
Likewise, according to \citet{em19}, E-PI{\textquoteright}s maximum tangential wind at the surface exceeds the maximum tangential wind at the boundary layer top. This provides a complementary perspective on the discussed incompatibility between E-PI{\textquoteright}s assumptions at $\pt v/\pt z = 0$.
The thermal wind equation and the assumption of slantwise neutrality constrain the slope of the angular momentum surfaces \citep[see][their Eq.~(A.14)]{stern2009}. This predicted slope is never vertical (unless $r = 0$), although it must be so where $\pt v/\pt z = 0$. 

If the $M$ and $s^*$ surfaces are vertical within the boundary layer, and then on the top of the boundary layer 
their slope abruptly changes to the one constrained by E-PI{\textquoteright}s Eq.~\eqref{em12} for the free troposphere, then the
radial gradients of $M$, $s^*$, $T$ and $p$ will all have infinite derivatives over $z$ on the top of the boundary layer where this discontinuity occurs. 
According to \citet{emanuel11}, E-PI{\textquoteright}s boundary layer closure requires $s^*$ to be well mixed along
the approximately vertical $M$ surfaces. The condition $\pt s^*/\pt z = 0$ is generally incompatible with  $\pt^2 s^*/(\pt z \pt r) \ne 0$ resulting from the discontinuity on the boundary layer top ($\pt s^*/\pt z$ can be zero only at a certain radius where it must change sign over $r$).   It is therefore pertinent to check to which degree the approximation of verticality is essential for E-PI{\textquoteright}s boundary layer closure.

E-PI{\textquoteright}s boundary layer closure constrains $ds^*/dM = (\pt s^*/\pt M)_z$ at the top of the boundary layer as the ratio of the vertical fluxes 
$F_{s^*}$ and $F_M$ of $s^*$ and $M$ at the sea surface. Originally, \citet[][p.~593]{em86} applied his Eq.~(27), valid {\textquotedblleft}for any conservative variable $c$ assumed to be well-mixed in the vertical  within a turbulent boundary layer{\textquotedblright} to angular momentum $M$, but noted at the same time that $M$ {\textquotedblleft}may not be well mixed in the vertical{\textquotedblright}.\footnote{For example, in \citeauthor{bryan09b}{\textquoteright}s~\citeyearpar{bryan09b} control simulation designed to check the E-PI assumptions, $M$ contours near the surface are approximately horizontal (see their Fig.~4).} 

For steady axisymmetric flow, the conservation equation for $c$ reads
\beq\label{cons}
\frac{1}{r} \frac{\pt (c r u)}{\pt r} + \frac{\pt (c w)}{\pt z} = \sigma,
\eeq
where $\sigma$ is the source/sink, which can also account for diffusion processes \citep[e.g.,][their Eq.~(13)]{bryan09a}.
Assuming, following \citet[][p.~593]{em86}, that within the boundary layer $c = \rho$ is approximately constant, we have
\beq\label{cons2}
\frac{1}{r} \frac{\pt (r u)}{\pt r} + \frac{\pt w}{\pt z} = 0.
\eeq
Taking into account Eq.~\eqref{cons2} and neglecting the horizontal diffusion, for  $c = M$, Eq.~\eqref{cons} becomes
\beq\label{consM}
u \frac{\pt M}{\pt r} + w \frac{\pt M}{\pt z} = - \frac{\pt F_M}{\pt z},
\eeq
where $F_M$ is the vertical flux of $M$.  In \citeauthor{em86}{\textquoteright}s~\citeyearpar{em86} 
Eq.~(27), the second term on the left-hand side of Eq.~\eqref{consM} is absent. The error introduced by omitting the vertical term depends 
on the ratio $w/u$. If, near the point of maximum wind, $u \to 0$, while $w$ is at its maximum \citep[][p.~3054]{bryan09b}, the error
can be large even if $\pt M/\pt z$ is relatively small compared to $\pt M/\pt r$. On a related note, the absolute angle between
the $M$ and $s^*$ surfaces does not provide information about how well $s^* = s^*(M)$ is satisfied in the E-PI context \citep[cf.][their~Fig.~3b,d,f]{tao2019}.

In the general case of $\pt M/\pt z \ne 0$ and $\pt s^*/\pt z \ne 0$, E-PI{\textquoteright}s boundary layer closure requires $s^*=s^*(M)$ within the boundary layer. Writing Eq.~\eqref{cons} for $c = s^*$ and using $\pt s^*/\pt r = (ds^*/dM) \pt M/\pt r$ and $\pt s^*/\pt z = (ds^*/dM) \pt M/\pt z$,
and Eq.~\eqref{consM}, we obtain $ds^*/dM = (\pt F_{s^*}/\pt z)/(\pt F_{M}/\pt z)$. If $ds^*/dM$ is constant and if both fluxes become zero at the top of
the boundary layer, this can be integrated over $z$ to yield $ds^*/dM = F_{s^*}/F_{M}$. But $s^*=s^*(M)$ within the boundary layer cannot be justified due to turbulence.

\citet[][p.~2239]{emanuel11} referred to the study of \citet{bryan09b} as demonstrating that E-PI{\textquoteright}s boundary closure {\textquotedblleft}is well satisfied in axisymmetric numerical simulations{\textquotedblright}. However, in the  control simulation of \citet[][p.~3049]{bryan09b}, E-PI{\textquoteright}s boundary layer closure at the radius of maximum wind is violated by $50\%$: the diagnosed ratio of surface fluxes is $1.5$-fold greater than the diagnosed $(\pt s^*/\pt M)_z$ at the top of the boundary layer, as shown in \citeauthor{bryan09b}{\textquoteright}s~\citeyearpar{bryan09b} Fig.~6. This discrepancy  is smaller in their Fig.~7, which presents the simulations of \citet{bryan09a}. But those simulations were made with a different parameter $l_v$ that controls vertical turbulence effects ($l_v = 100$~m in the control simulation of \citet[][their Fig.~6]{bryan09b} and $l_v = 200$~m for simulations of \citet{bryan09a} shown in   \citeauthor{bryan09b}{\textquoteright}s~\citeyearpar{bryan09b} Fig.~7). 

Importantly, according to \citet[][see their Fig.~2]{bryan09a}, the value of $l_v$ does not influence the maximum wind speed. At the same time, as the comparison of  \citeauthor{bryan09b}{\textquoteright}s~\citeyearpar{bryan09b}  Figs. 6 and 7 suggests, parameter $l_v$ is instrumental in bringing E-PI{\textquoteright}s boundary layer closure in agreement with the simulations.  If there exist model parameters that control whether E-PI{\textquoteright}s boundary layer closure is satisfied, and if such parameters make no impact on the maximum intensity, the inference is that the  maximum intensity may not be as profoundly dependent on local surface fluxes as E-PI presumes. This requires further clarifications.

\section{Discussion and conclusions}
\label{dis}

We applied E-PI{\textquoteright}s assumptions to the altitude of maximum tangential wind ($\pt v/\pt z = 0$), which, according to observations
and numerical models, is located near the top of the boundary layer. We showed that here E-PI{\textquoteright}s assumptions are mutually incompatible
and only allow for a trivial solution $\pt s^*/\pt r = 0$ and $\pt M/\pt r = 0$. We also applied E-PI{\textquoteright}s assumptions to the point of maximum tangential wind ($\pt v/\pt z = \pt v/\pt r = 0$) and showed that here their mutual incompatibility results in contrasting predictions concerning the radial temperature gradient. The thermal wind equation requires it to be positive, while E-PI{\textquoteright}s key equation constrains it to be negative and dependent 
on the outflow temperature.

E-PI is based on merging the free troposphere constraints with the boundary layer constraints. The incompatibility
of its assumptions pertains to the altitude of maximum tangential wind located on the border between the two atmospheric layers,
and has implications for both. We have shown that at the altitude of maximum tangential wind the flow must be supergradient and that its
relaxation to the gradient-wind balance in the free troposphere disturbs the constancy of $ds^*/dM$ required by E-PI.
In the boundary layer, the verticality of $M$ surfaces assumed in E-PI from the sea surface up to the boundary layer top, is not compatible with the non-verticality of $M$ surfaces required by E-PI in the free troposphere. This disturbs the relationship between
$ds^*/dM$ on the boundary layer top and the ratio of the surface fluxes of $s^*$ and $M$ that is required for E-PI{\textquoteright}s boundary layer closure.

Without addressing these theoretical issues, continued efforts to verify E-PI, or its elements, with numerical simulations
may not be conclusive regarding the general validity of E-PI. Increasing model complexity
without a matching increase in the quality of its independent constraints, leads to fuzzier conclusions \citep{puy2022}.
In such a situation, the results of numerical simulations can be misleading. We discuss one example below. 

One of our reviewers referred to the study of \citet[][]{tao2019} as demonstrating an agreement between model simulations and 
E-PI{\textquoteright}s Eq.~\eqref{em12}, from which Eq.~\eqref{em13} derives. If $(T_b - T_o)$, $ds^*/dM$ and $M$ are known, $v^2 \simeq M^2/r^2$ can be diagnosed from Eq.~\eqref{em12}, see  \citeauthor{tao2019}{\textquoteright}s~\citeyearpar{tao2019}  Eq.~(3):
\beq\label{tao3}
v^2 = -(T_b - T_o) M \frac{ds^*}{dM}.
\eeq
However, while Fig.~5c,f,i of \citet{tao2019} does indeed display a good agreement between predicted and model-derived velocities,
it does not validate E-PI{\textquoteright}s Eq.~\eqref{em12}. 
 First, we note that \citet[][p.~3007 and p.~2999]{tao2019} recognized that {\textquotedblleft}the model simulated flow can be supergradient within the boundary layer{\textquotedblright} and thus  {\it defined}  the top of the boundary layer as {\textquotedblleft}the altitudes where the maximum tangential winds first agree quantitatively with the maximum gradient winds{\textquotedblright}, i.e., deliberately choosing the altitudes where the gradient-wind balance (approximately) holds. Accordingly, $V_t$ in \citeauthor{tao2019}{\textquoteright}s~\citeyearpar{tao2019}  Fig.~5c,f,i  is not the actual modeled maximum tangential wind, but  {\textquotedblleft}the modeled maximum tangential wind at the boundary layer top{\textquotedblright} thus defined.  \citet[][p.~3007]{tao2019} emphasized that in all their simulations the diagnosed maximum $v$ from Eq.~\eqref{tao3} (shown by red dots in their Fig.~5c,f,i) was {\textquotedblleft}consistently smaller{\textquotedblright} than the actual modeled maximum tangential wind (not reported).

From this one could conclude that E-PI{\textquoteright}s Eq.~\eqref{em12} could be valid if not at the point of maximum tangential wind but 
at least at a certain altitude where the gradient-wind balance (approximately) holds. But there is an additional caveat.
\citet[][p.~2999]{tao2019} correctly noted, see Eq.~\eqref{em11} above, that in E-PI $ds^*/dM$ should be constant on $M$ surfaces 
because of the assumed congruence of $s^*$ and $M$ surfaces.  Thus, \citet[][p.~2999 and their Fig.~4]{tao2019} diagnosed $ds^*/dM$ {\it not at the same point} where they diagnosed the tangential wind (at the boundary layer top), but in the outflow region in the upper troposphere. However, their own Fig.~3b,d,f makes it clear that $s^*$ and $M$ surfaces are not congruent over much  of the eyewall (between approximately $2$~and $9$~km). This means that the values of $ds^*/dM$ in the boundary layer and in the outflow region cannot be assumed equal, as $ds^*/dM$ varies where $s^* \ne s^*(M)$.

While \citet{tao2019} did not analyze how $ds^*/dM$ varied along the $M$ surfaces, other studies indicate that such variation can be substantial.
Figure~5b of \citet{peng2018} shows simultaneously $s^*$ and $M$ contours and readily allows for the estimation of $ds^*/dM$ in their modeled steady-state vortex. In the boundary layer at the radius $r = 30$~km of maximum wind  there are three $M$ contour intervals in one $s^*$ contour interval, while in the outflow region at $r = 80$~km  there are only two. This indicates a $1.5$-fold increase of the absolute value of $ds^*/dM$. This increase displays a tendency to continue  at larger radii in the outflow (not shown in \citeauthor{peng2018}{\textquoteright}s~\citeyearpar{peng2018}  Fig.~5b). This is consistent with the decline in $|ds^*/dM|$ for supergradient wind discussed in Section~\ref{key}. If a similar pattern holds for the simulations of \citet{tao2019}, then their use of $ds^*/dM$ from the outflow region would lead to an overestimate of $v^2$ as diagnosed from Eq.~\eqref{em12} by a factor of $1.5$ or more \citep[cf.][]{peng2018}.

In summary, we are not aware of any studies, either observational or modelling, where the validity of E-PI{\textquoteright}s Eqs.~\eqref{em12} and \eqref{em13} would be demonstrated together with the validity of their underlying assumptions. Our alternative Eq.~\eqref{alt1}  suggests that if a constraint on $\mathcal{C}$ (the degree of adiabaticity of the radial temperature gradient) at the point of maximum wind is found, one could dispense with the consideration of the upper troposphere -- as E-PI dispenses with the consideration of boundary layer dynamics when constraining $\mathcal{C}$  from the free troposphere considerations alone (see Eq.~\eqref{comp1}). Considering the boundary layer dynamics, \citet{mn21} suggested that $\pt T/\pt r = 0$ and $\mathcal{C} = 1$ at the radius of maximum wind is a plausible assumption, which accuracy could be further investigated.

However, from our perspective, the main, and fundamental, problems of E-PI (and of any other local approach, including the alternative Eq.~\eqref{alt1}) pertain to the boundary layer closure.
Some were discussed here, but see also \citet{mn21}. Storm intensity is an integral property of the entire storm’s energetics, whereby the energy released over a large area is concentrated in the eyewall to generate maximum  wind. It cannot be a local function of the highly variable heat input at the radius of maximum wind (even if one could tune a model to suggest otherwise).  We argue for a principally different approach to storm dynamics.

\section*{Acknowledgments}
The authors are grateful to three reviewers for their useful comments. Our response to the reviewers can be found in appendix~\ref{respr}. Work of A.M. Makarieva is partially funded by the Federal Ministry of Education and Research (BMBF) and the Free State of Bavaria under the Excellence Strategy of the Federal Government and the L\"ander, as well as by the Technical University of Munich -- Institute for Advanced Study. The authors would like to thank V\'{a}clav Vacek, Jan Pokorn\'{y} and  Milan Vlach for stimulating discussions and support.

\appendix
\setcounter{section}{0}%
\section{How does the gradient wind change over $z$?}
\label{prgr}

\setcounter{equation}{0}%
\renewcommand{\theequation}{A\arabic{equation}}%

From Eq.~\eqref{he} we have $\pt p/\pt z =-p/h$, where $h \equiv RT/(g M_a)$, while $\alpha = gh/p$. Using these relations we have:
\begin{align}
\frac{\pt}{\pt z} \left(\frac{h}{p} \frac{\pt p}{\pt r}\right) &= \frac{\pt p}{\pt r} \frac{\pt }{\pt z} \left(\frac{h}{p}\right)- \frac{h}{p} \frac{\pt }{\pt r} \left(\frac{p}{h\vphantom{p}}\right)=
\frac{\pt p}{\pt r} \left(\frac{1}{p} \frac{\pt h}{\pt z} - \frac{h}{p^2}\frac{\pt p}{\pt z} \right) -\frac{h}{p} \left(\frac{1}{h} \frac{\pt p}{\pt r} -
\frac{p}{h^2} \frac{\pt h}{\pt r} \right) = \nonumber\\ \label{prgr1}
&= \frac{1}{p} \frac{\pt p}{\pt r} \frac{\pt h}{\pt z} + \frac{1}{h} \frac{\pt h}{\pt r} =\frac{1}{p} \frac{\pt p}{\pt r} \left(\frac{\pt h}{\pt p}\right)_r \frac{\pt p}{\pt z}
+ \frac{1}{h}\left(\frac{\pt h}{\pt p}\right)_z \frac{\pt p}{\pt r} = (K-1)\frac{1}{h}\left(\frac{\pt h}{\pt p}\right)_r \frac{\pt p}{\pt r} ,
\end{align}
where $K \equiv (\pt h/\pt p)_z/(\pt h/\pt p)_r = (\pt T/\pt p)_z/(\pt T/\pt p)_r$. Note that when $(\pt T/\pt p)_r = \Gamma$, then $K = 1 - \mathcal{C}$, see Eq.~\eqref{C}.

The relative change of gradient wind with height is given by
\begin{equation}\label{prgr3}
\left(\alpha \frac{\pt p}{\pt r}\right)^{-1}\frac{\pt}{\pt z} \left(\alpha \frac{\pt p}{\pt r}\right)  =  (K-1)\frac{p}{h^2} \left(\frac{\pt h}{\pt p}\right)_r = (1-K)\frac{\pt \ln h}{\pt z} =(1-K)\frac{\pt \ln T}{\pt z} = - \frac{\pt \ln b}{\pt z}.
\eeq
The last equality in Eq.~\eqref{prgr3} follows from Eq.~\eqref{bz} under assumption that $\pt v/\pt z = 0$; it quantifies the change in $b$ due to thermodynamics.

With $\pt v/\pt z = 0$, the supergradiency factor $b$ remains constant over $z$ if  the horizontal and vertical gradients of temperature over pressure are equal ($K=1$). If the atmosphere is horizontally isothermal ($K = 0$), then, with a moist adiabatic lapse rate $-\pt T/\pt z \simeq 5$~K~km$^{-1}$, the relative increase in $b$ will be under $2\%$ over $1$~km. If the horizontal temperature lapse rate is minus moist adiabatic ($K \simeq  -1$,  $\mathcal{C}\simeq 2$), as E-PI approximately requires (see Section~\ref{constr0}), then $b$ will increase by no more than $4\%$ over $1$~km.

\setcounter{equation}{0}%
\renewcommand{\theequation}{B\arabic{equation}}%

\section{Deriving an alternative for E-PI{\textquoteright}s Eq.~\eqref{em13}}
\label{altf}

Moist entropy $s$ per unit mass of dry air is defined as (e.g., Eq.~(2) of \citet[][]{em88},  Eq.~(A4) of  \citet[][]{pa11})
\begin{equation}\label{s}
s = (c_{pd} + q_t c_l) \ln \frac{T}{T'} - \frac{R}{M_d} \ln \frac{p_d}{p'} + q \frac{L_v}{T} - q\frac{R}{M_v}\ln \mathcal{H}.
\end{equation}
Here,  $L_v = L_{v}(T') + (c_{pv} - c_l)(T-T')$ is the latent heat of vaporization (J~kg$^{-1}$); 
$q \equiv \rho_v/\rho_d \equiv \mathcal{H}q^*$ is the water vapor mixing ratio; $\rho_v$ is water vapor density; $\mathcal{H}$ is relative humidity;  $q^* = \rho_v^*/\rho_d$,  $q_l = \rho_l/\rho_d$, and  $q_t = q + q_l$ are the mixing ratio for saturated water vapor, liquid water, and total water, respectively; $\rho_d$, $\rho_v^*$, and $\rho_l$ are the density of dry air, saturated water vapor and liquid water, respectively;  $c_{pd}$ and $c_{pv}$ are the specific heat capacities of dry air and water vapor at constant pressure; $c_l$ is the specific heat capacity of liquid water; $R = 8.3$~J~mol$^{-1}$~K$^{-1}$ is the universal gas constant; $M_d$ and $M_v$ are the molar masses of dry air and water vapor,
respectively;  $p_d$ is the partial pressure of dry air; $T$ is the temperature; $p'$ and $T'$ are reference air pressure and temperature.

For saturated moist entropy $s^*$ ($q = q^*$, $\mathcal{H} = 1$) we have
\begin{gather} \label{Tds2}
Tds^* = (c_{pd} + q_t c_l)dT  - \frac{R T}{M_d} \frac{dp_d}{p_d} + L_v dq^* +q^* dL_v  - q^* L_v \frac{dT}{T}=\left(c_p - \frac{q^* L_v}{T}\right)dT - \frac{R T}{M_d} \frac{dp_d}{p_d} + L_v dq^*  , 
\end{gather}
where $c_p  \equiv c_{pd} + q^* c_{pv} + q_l c_l$ and $dL_v =(c_{pv} - c_l) dT$.  Equation \eqref{Tds2} additionally assumes $q_t = \mathrm{const}$ (reversible adiabat).

The ideal gas law for the partial pressure $p_v$ of  water vapor is
\begin{equation}\label{igv}
p_v = N_v RT,  \quad N_v = \frac{\rho_v}{M_v} , 
\end{equation}
where  $M_v$ and $\rho_v$ are the molar mass and density of water vapor. Using Eq.~(\ref{igv}) with $p_v = p_v^*$ in the
definition of $q^*$ 
\begin{equation}\label{q}
q^* \equiv \frac{\rho_v^*}{\rho_d} = \frac{M_v}{M_d} \frac{p_v^*}{p_d} \equiv \frac{M_v}{M_d} \gamma_d^*,\quad 
\gamma_d^* \equiv\frac{p_v^*}{p_d},
\end{equation}
and  applying the Clausius-Clapeyron law 
\begin{equation}\label{CC}
\frac{dp_v^*}{p_v^*} = \frac{L}{RT} \frac{dT}{T}, \quad L \equiv L_v M_v , 
\end{equation}
we obtain for the last term in Eq.~(\ref{Tds2})
\begin{equation}\label{dq}
L_vdq^* = L_v\frac{M_v}{M_d}\left(\frac{dp_v^*}{p_d} - \frac{p_v^*}{p_d} \frac{dp_d}{p_d}\right)=L_v\frac{M_v}{M_d}\left(\frac{p_v^*}{p_d}\frac{dp_v^*}{p_v^*} - \frac{p_v^*}{p_d} \frac{dp_d}{p_d}\right)
=L_vq^*\left(\frac{L}{RT}\frac{dT}{T} -\frac{dp_d}{p_d}\right) . 
\end{equation}

Using the Clausius-Clapeyron law (\ref{CC}), the ideal gas law $p_d = N_d RT$, where $N_d = \rho_d /M_d$, and noting that $p= p_v^* + p_d$, we obtain for the last but one term in Eq.~(\ref{Tds2})
\begin{equation}\label{pd}
\frac{RT}{M_d}\frac{dp_d}{p_d}=\frac{RT}{M_d} \left(\frac{dp}{p_d} - \frac{dp_v^*}{p_d}\right)  = \frac{dp}{M_dN_d} - \frac{RT p_v^*}{M_d p_d}\frac{dp_v^*}{p_v^*}
=\frac{dp}{\rho_d} - L_v \frac{M_v p_v^*}{M_d p_d}\frac{dT}{T} = \frac{dp}{\rho_d} - q^*L_v\frac{dT}{T}. 
\end{equation}

Taking into account  Eq.~\eqref{pd}, Eq.~\eqref{Tds2} reads
\begin{gather} \label{Tds2ss}
Tds^* =c_p dT - \alpha_d dp + L_v dq^*  . 
\end{gather}
Putting Eqs.~(\ref{dq})  into Eq.~(\ref{Tds2ss}) yields
\begin{gather}\label{Tds3}
Tds^* =\left(c_p +  \frac{L_v q^*}{T} \frac{L(1 + \gamma_d^*)}{RT}\right)dT - \left(1 + \frac{L \gamma_d^*}{RT} \right) \frac{dp}{\rho_d} = 
-(1 +\zeta) \alpha_d \left( 1 - \frac{1}{\Gamma}\frac{dT}{dp} \right) dp .
\end{gather}
Here
\begin{equation}\label{Gm}
\Gamma \equiv \frac{\alpha_d}{c_p} \frac{1 + \zeta}{1 +  \mu\zeta (\xi + \zeta)},\quad 
\xi \equiv \frac{L}{RT}, \quad \zeta \equiv \xi \gamma_d^* \equiv \frac{L}{RT}\frac{p_v^*}{p_d} \equiv \frac{L_v q^*}{\alpha_d p_d}, \quad \mu  \equiv \frac{R}{C_p}= \frac{2}{7} , 
\end{equation}
where $\alpha_d \equiv 1/\rho_d$ is the volume per unit mass of dry air and $C_p \simeq c_p M_d$ is the molar heat capacity of air at constant pressure.

\section{Response to the reviewers}
\label{respr}

{\color{blue} \it
\noindent
Jan 19, 2023

\vspace{0.15cm}
\noindent
Ref.: JAS-D-22-0144

\vspace{0.15cm}
\noindent
Editor Decision

\vspace{0.45cm}\noindent
Dear Dr. Makarieva,

\vspace{0.45cm}\noindent
I am now in receipt of three reviews of your revised manuscript.  All three reviewers are generally sa\-tis\-fied by the changes you have made in your revisions, and Reviewer 1 now recommends acceptance.  Reviewers 2 and 3 both have a few very minor comments, and so to give you the opportunity to make any additional changes to address these suggestions, my decision is Minor Revision. The reviews are enclosed below.

\vspace{0.15cm}\noindent
We invite you to submit a revised paper by Mar 20, 2023. If you anticipate problems meeting this deadline, please contact me as soon as possible at  \underline{Stern.JAS@ametsoc.org}.

\vspace{0.15cm}\noindent
Along with your revision, please upload a point-by-point response that satisfactorily addresses the concerns and suggestions of each reviewer. To help the reviewers and Editor assess your revisions, our journal recommends that you cut-and-paste the reviewer and Editor comments into a new document. As you would conduct a dialog with someone else, insert your responses in a different font, different font style, or different color after each comment. If you have made a change to the manuscript, please indicate where in the manuscript the change has been made.  (Indicating the line number where the change has been made would be one way, but is not the only way.)

\vspace{0.15cm}\noindent
Although our journal does not require it, you may wish to include a tracked-changes version of your manuscript. You will be able to upload this as {\textquotedblleft}additional material for reviewer reference{\textquotedblright}. Should you disagree with any of the proposed revisions, you will have the opportunity to explain your rationale in your response.

\vspace{0.15cm}\noindent
Please go to  \underline{www.ametsoc.org/PUBSrevisions} and read the AMS Guidelines for Revisions. Be sure to meet all recommendations when revising your paper to ensure the quickest processing time possible.

\vspace{0.15cm}\noindent
When you are ready to submit your revision, go to  \underline{https://www.editorialmanager.com/amsjas/} and log in as an Author. Click on the menu item labeled {\textquotedblleft}Submissions Needing Revision{\textquotedblright} and follow the directions for submitting the file.

\vspace{0.15cm}\noindent
Thank you for submitting your manuscript to the Journal of the Atmospheric Sciences. I look forward to receiving your revision.

\vspace{0.45cm}\noindent
Sincerely,

\vspace{0.35cm}
\noindent
Dr. Daniel Stern

\noindent
Editor

\noindent
Journal of the Atmospheric Sciences
}

\clearpage
\vspace{0.4cm}
\noindent
Jan 30, 2023

\vspace{0.15cm}
\noindent
Ref.: JAS-D-22-0144

\vspace{0.15cm}
\noindent
Submission of revised manuscript

\vspace{0.45cm}
\noindent
Dear Dr. Stern,

\vspace{0.15cm}
\noindent
Thank you for consideration of our work. We have implemented the minor revisions as advised
by the reviewers. We are grateful to the reviewers and the Editor for constructive suggestions
that helped us clarify and develop our arguments.
We look forward to hearing from you in due course.

\vspace{0.45cm}
\noindent
Yours sincerely,

\vspace{0.35cm}
\noindent
Anastassia Makarieva

\vspace{0.15cm}
\noindent
Andrei Nefiodov

\newpage

\vspace{0.4cm}
{\color{blue} \it
\noindent
\begin{center}\textbf{REVIEWER COMMENTS} \end{center}

\vspace{0.15cm}
\begin{center}
\textbf{Reviewer \#1:}
\end{center} 

\vspace{0.15cm}\noindent
The authors have satisfactorily addressed all of my major and minor comments. I now recommend acceptance.

\vspace{0.15cm}\noindent 
{\color{black}\rm	
We are sincerely grateful to all our reviewers for their time and efforts. We have appreciated the constructive comments 
 and suggestions.}

\vspace{0.45cm}
\begin{center}
\textbf{Reviewer \#2:}
\end{center}

\vspace{0.15cm}\noindent
E-PI is not perfect. I agree especially with the problem in the boundary layer closure, but I also think we should appreciate its merit in connecting the whole troposphere and partially solving the {\textquoteleft}maximum potential intensity{\textquoteright} in a relative sense (given the environment variables, we are able to evaluate the potential). The thermal wind assumptions hold pretty well (not saying $100\%$ match) in the free troposphere above the boundary layer, and equation (11) is a good approximation for places the assumptions apply, in my viewpoint, the main issue with E-PI is that it extended equation (11) too far down to the location of the maximum wind and used a simple slab boundary layer closure. The imbalance in the boundary layer does impact the maximum wind, then how can we incorporate it in the derivations?

\vspace{0.15cm}\noindent 
{\color{black}\rm	
While the current research focus is on superintensity, we believe that it could be informative to
identify reasons why most real tropical storms have their intensities well below E-PI.  We have suggested that the entropy difference across the air-sea interface grows with increasing radial velocity -- i.e., it is a function of the secondary circulation \citep{mn21}. Here we have shown that supergradiency at the point of maximum wind is also a function of radial velocity and its spatial change. It could be informative to assess in theory what kind of variability the combination of these factors should impart to maximum intensity if the latter were determined by local heat flux at the radius of maximum wind, and then to compare this predicted variability to observations.}

\vspace{0.15cm}\noindent
Minor comments:

\vspace{0.15cm}\noindent
L30: Suggest to add a transition sentence like {\textquoteleft}Studies have shown that the maximum wind is larger than E-PI due to supergradient wind...{\textquoteright} from E-PI to {\textquoteleft}Here we re-consider...{\textquoteright}

\vspace{0.15cm}\noindent 
{\color{black}\rm	
Revised as suggested.}

\vspace{0.15cm}\noindent
L76: in the left-hand side --$>$ on the left-hand side

\vspace{0.15cm}\noindent 
{\color{black}\rm	
Done.}

\vspace{0.15cm}\noindent
L111: add {\textquoteleft}following M or s* surface{\textquoteright} at the end

\vspace{0.15cm}\noindent 
{\color{black}\rm	
Revised as suggested.}

\vspace{0.45cm}
\begin{center}
\textbf{Reviewer \#3:}
\end{center}

\vspace{0.15cm}\noindent
I appreciate the authors for the informative revisions. I only have two minor comments, which are related to two technical details that I struggle to understand.

\vspace{0.15cm}\noindent
1) L210-212 Could you add a brief reasoning for the conclusion {\textquotedblleft}it follows that the secondary derivative of u ... is zero...{\textquotedblright} ?

\vspace{0.15cm}\noindent 
{\color{black}\rm
We added an additional footnote (\# 4) for clarification. Since $\partial u/\partial z = 0$ at the point where $\partial v/\partial z = 0$, it is by definition a \textit{stationary point}. There are three types of stationary points: local maximum, local minimum and stationary inflection. If the first derivative changes sign (from positive to negative, or vice versa) at that point, the stationary point is a turning point (maximum or minimum). Inflection point is also a  stationary point, but it is not a turning point. More precisely, here we deal with a rising point of inflection, because the first derivative of the function is not negative in the vicinity of the stationary point ($\partial u/\partial z \ge 0$). As it is a point of inflection, we have $\partial^2 u/\partial z^2 = 0$ at this point.
}

\vspace{0.15cm}\noindent
2) L392-L394 Could you mention which equation you referred to so that the last equality of Eq. (A2) is valid?

\vspace{0.15cm}\noindent 
{\color{black}	\rm
The last equality in Eq.~(A2) follows from Eq.~(25) under assumption that $\partial v/\partial z = 0$. We have clarified this in the sentence following Eq.~(A2).

}

{\color{black}	\rm


}

\end{document}